\documentstyle[prd,aps,floats]{revtex}

\begin{document}


\draft

\twocolumn[\hsize\textwidth\columnwidth\hsize\csname
@twocolumnfalse\endcsname

\title{
Cosmological constant, dilaton field and Freund-Rubin
compactification
}

\author{
Takashi Torii$^{(a)}$ and  
Tetsuya Shiromizu$^{(b,a)}$
}

\address{ $^{(a)}$
Advanced Research Institute for Science and Engineering,
Waseda University, Tokyo 169-8555, Japan
}

\address{$^{(b)}$
Department of Physics, Tokyo Institute of Technology,
Tokyo 152-8551, Japan
}

\date{\today}

\maketitle

\begin{abstract}
We discuss Freund-Rubin compactification with a
cosmological constant and the
dilaton field, and examine the stability of
spacetime at low energy.
Minkowski or de Sitter spacetime can be obtained if
the dilation field is turned off, but only anti-de Sitter
spacetime is realized in the presence of the
dilaton field.
The stability of spacetime depends on the dimensionality
of spacetime and the compactified space, and the
coupling constant of the dilaton field $a$.
In the $a=1$ case, which corresponds to
superstring theories, the anti-de Sitter vacuum is
stable at least in the linear level.
\end{abstract}



\vskip2pc]

\setcounter{footnote}{1}
\renewcommand{\thefootnote}{\fnsymbol{footnote}}


Compactification from higher dimensions to
four dimensions and
stabilization of the compactified space
is a long-standing problem. It is well known that
Freund-Rubin (FR) compactification\cite{FR} naturally
introduced from supergravity
gives us stable compact dimensions
via a balance between the form field and the curvature
at the classical level.
Spacetime after
compactification will becomes anti-de Sitter (adS) spacetime.
Recently the realization of $M_p \times
S^q$, where $M_p$ is the $p$-dimensional Minkowski,
de Sitter or adS
spacetime, has been discussed, introducing an
additional cosmological constant to
the total action\cite{Wald,Bousso}. The subject of stability
has also been addressed and some parameter windows open for 
stable
de Sitter or Minkowski as well as adS spacetimes.
This is an impressive
result, although the new ingredient is just
the additional cosmological constant.
We would like to point out, however, that a stable background with
Minkowski or de Sitter spacetime is not
permitted if the (NS) dilaton coupling exists.
It is only adS spacetime that is permitted
as the background solution.
We briefly address the stability of the obtained adS
vacuum in what follows.


We begin with the $D\;(=p+q)$-dimensional action in the
string frame:
%
\begin{eqnarray}
S&=&\frac{1}{16\pi} \int d^{p+q}x \sqrt {-\tilde{G}}
\left[e^{-2a\Phi}
\Bigl( \tilde{R}-2\Lambda+4\partial_M\Phi \partial^M\Phi \Bigr)
\right.
\nonumber
\\
&& \;\;\;\;\;\;\;\;\;\;
\left. -\frac{1}{2q!}F_q^2 \right], \label{action}
\end{eqnarray}
%
where $\Lambda$ is a positive cosmological constant and
$F_q$ is the $q$-form field.
$\Phi$ is  the  (NS) dilaton field and we leave its coupling
constant $a$ as a free parameter. Here we do not ask about the 
origin of $\Lambda$.

It is convenient to transform the action to the Einstein
frame by the following conformal transformation
%
\begin{eqnarray}
\tilde{G}_{MN}=\Omega^2 G_{MN},
\;\;\;\;\;
\Omega^2 = e^{4a\Phi/(D-2)}.
\end{eqnarray}
%
Then the action in the Einstein frame becomes
%
\begin{eqnarray}
S &=&\frac{1}{16\pi} \int d^{p+q}x {\sqrt {-G}}
\left[R-2e^{\lambda_c\Psi}\Lambda
-\frac12 \partial_M\Psi \partial^M\Psi
\right. \nonumber
\\
&& \;\;\;\;\;\;\;\;\;\;\;\;
\left.
-\frac{1}{2q!}e^{\lambda_q\Psi}F_q^2\right],
\label{einframe}
\end{eqnarray}
%
where
%
\begin{eqnarray}
\Psi &:=& \mu\Phi,
\\
\mu &:=& \sqrt{8\left[\frac{(D-1)a^2}{D-2}-1\right]},
\\
\lambda_c &:= &\frac{4a}{(D-2)\mu},
\\
\lambda_q &:= &\frac{2a(p-q)}{(D-2)\mu}.
\end{eqnarray}
%
In superstring theories, $D=10$ and $a=1$ and we find
$\mu=1$,  $\lambda_c=1/2$ and
$\lambda_q=(10-2q)/5$\cite{quanta}.
Note that the coupling constant of the dilaton field must satisfy
$a^2>(D-2)/(D-1)$.

We employ the following metric form
%
\begin{eqnarray}
\label{met}
ds^2=g_{\mu\nu}(x)dx^\mu dx^\nu + e^{2\phi(x)}d\Omega_q^2.
\end{eqnarray}
%
Since $\phi$ does not depend on the internal coordinate, 
this metric describes the homogeneous mode of the internal
space.
In this case the form field equation gives the solution
%
\begin{eqnarray}
F_q=c *_q 1,
\end{eqnarray}
%
and the dilaton field  equation becomes
%
\begin{eqnarray}
& & \nabla_{\mu} \nabla^{\mu} \Psi
+q (\nabla_{\mu} \phi)(\nabla^{\mu} \Psi)
=\frac{\partial V}{\partial \Psi},
\end{eqnarray}
%
where $\ast_q$ is the Hodge dual in the $q$-dimensional space and
the potential $V$ is
%
\begin{eqnarray}
V=2e^{\lambda_c\Psi}\Lambda
+\frac{c^2}{2} e^{-2q\phi} e^{\lambda_q\Psi}. \label{pot0}
\end{eqnarray}
%

The $D$-dimensional Einstein equations become
%
\begin{eqnarray}
R_{\mu \nu} &=&
\frac12\partial_{\mu} \Psi \partial_{\nu} \Psi
\nonumber
\\
&& \;
+\frac{2}{D-2}\left[e^{\lambda_c \Psi}\Lambda
-\frac{(q-1)c^2}{4}e^{-2q \phi}e^{\lambda_q \Psi}
\right]g_{\mu \nu},
\nonumber
\\
\label{ein1}
\end{eqnarray}
%
\begin{eqnarray}
R_{ij} &=&
\frac{2}{D-2}\left[e^{\lambda_c \Psi}\Lambda
+\frac{(p-1)c^2}{4}e^{-2q \phi}e^{\lambda_q \Psi}
\right]g_{ij}.
\nonumber
\\
\label{ein2}
\end{eqnarray}


When $\phi\equiv 0$, which corresponds to FR
compactification with the additional cosmological
constant\cite{Wald,Bousso}, the
Einstein equations may give
$M_p\times M_q$ where $M_p$ is de Sitter, Minkowski or adS
spacetime
and $M_q$ is sphere, flat space or hyperboloid,
depending on the values of $\Lambda$ and
$c$. Let us find the background solution.
Before discussing the dilatonic case, we briefly summarize the
non-dilatonic case for comparison.

The non-dilatonic case is realized by imposing $\Psi\equiv 0$ and
$\lambda_c=\lambda_q=0$. The Einstein equations (\ref{ein1}) and
(\ref{ein2}) become
%
\begin{eqnarray}
{}^{(p)}R_{\mu \nu} &=&
\frac{c^2}{2(D-2)}(f-f_p)g_{\mu \nu},
\end{eqnarray}
%
\begin{eqnarray}
{}^{(q)}R_{ij} &=&
\frac{c^2}{2(D-2)}(f-f_q)g_{ij},
\end{eqnarray}
where ${}^{(p)}R_{\mu \nu}$ and ${}^{(q)}R_{ij}$ are
$p$-  and $q$-dimensional Ricci curvature, respectively, and
$f=4\Lambda/c^2$, $f_p=q-1$ and $f_q=1-p$. Depending on the value
of $f$, the $p$-dimensional spacetime becomes de Sitter ($f>f_p$),
Minkowski ($f=f_p$) or adS ($f<f_p$) spacetime.
The adS case includes $\Lambda=0$,
which corresponds to
the original FR solution\cite{FR}.
On the other hand, the $q$-dimensional
space becomes
$S^q$ ($f>f_q$), $R^q$ ($f=f_q$) and $H^q$ ($f<f_q$). Hence the
$q$-dimensional space is no longer compact without
compactification by
suitable identifications in the latter two cases. See
Ref.~\cite{Bousso} for the stability analysis.


In the dilatonic case, the parameter regions of $\lambda_c$ and
$\lambda_q$
where   stationary points of the potential
function (\ref{pot0}) exist
are restricted.
When $\Lambda>0$, the constant dilaton field is realized for
$\lambda_c\lambda_q<0$,
which implies that $p<q$. The value of the dilaton field is
%
\begin{eqnarray}
\Psi \equiv\Psi_c &:=& \frac{1}{\lambda_c-\lambda_q}
\ln \left|\frac{c^2 \lambda_q}{4\lambda_c\Lambda}\right|
\nonumber \\
&=& \frac{1}{\lambda_c-\lambda_q}
\ln \left|\frac{(p-q)c^2}{8\Lambda}\right|.
\label{constdil}
\end{eqnarray}
%
This point is the minimum of the potential, independent of the
parameters.

When $\Lambda=0$, $\lambda_q=0$ (i.e. $p=q$) gives the constant
dilaton field. In this case, the
dilaton field is regarded as a massless scalar field
with the flat potential $V\equiv c^2/2$.
Hence it
takes an arbitrary constant value.

When $\Lambda<0$, the solution with a constant dilaton
field exists only when
$\lambda_c\lambda_q>0$, which implies that $p>q$ except for $p-q=2$.
The value of the dilaton field is also given by Eq.~(\ref{constdil}).
For $p-q=1$, the stationary point $\Psi_c$ is the local maximum,
while it is the local minimum for $p-q>2$.

The Einstein equations (\ref{ein1}) and (\ref{ein2}) become
%
\begin{eqnarray}
R_{\mu \nu}=
-\frac{p-1}{\ell_p^2}g_{\mu \nu},
\label{ein3}
\end{eqnarray}
%
\begin{eqnarray}
R_{ij} &=&
\frac{q-1}{\ell_q^2}g_{ij},
\label{ein4}
\end{eqnarray}
where the curvature radii are given by
%
\begin{eqnarray}
\frac{p-1}{\ell_p^2}& = & \frac{q-1}{\ell_q^2}
=\frac{1}{4}c^2e^{\lambda_q \Psi_c} \nonumber \\
&=& \left\{
\begin{array}{ll}
\frac{c^2}{4}
\left|\frac{(p-q)c^2}{8\Lambda}\right|^{-\frac{p-q}{p-q-2}}
& (p\ne q),
\\
\frac{c^2}{4} & (p=q)
\end{array}
\right.
\end{eqnarray}
Hence the background $p$-dimensional spacetime is always
adS and the $q$-dimensional space is sphere independently
of the cosmological constant.
It should also be noted that existence of the solution
depends on the sign of the
cosmological constant and dimensions $p$ and $q$. These are direct
effects of including the dilaton field.


Before we proceed to the stability analysis of the FR type
solutions obtained above, we derive the $p$-dimensional effective
theory to address their stability.
The $D$-dimensional action (\ref{einframe}) is
reduced to the  $p$-dimensional
action as
%
\begin{eqnarray}
S&=&\frac{1}{16\pi}\int d^p x \sqrt{-g}e^{q\phi}
\biggl[{}^{(p)}R +\frac{q(q-1)}{\ell_q^2}e^{-2\phi}
\nonumber
\\
&&\;\;\;\;\;
-\frac12 D_{\mu}\Psi D^{\mu}\Psi+q(q-1)D_{\mu}\phi D^{\mu}\phi
\nonumber
\\
&&\;\;\;\;\;
-2e^{\lambda_c \Psi}\Lambda
-\frac{c^2}{2}e^{\lambda_q \Psi}e^{-2q \phi}
\biggr].
\end{eqnarray}
Here we have used
%
\begin{eqnarray}
{}^{(q)}R=\frac{q(q-1)}{\ell_q^2},
\label{no}
\end{eqnarray}
and the assumption that $\phi$ is the homogeneous mode. 
It is derived from the form field equation 
$d(e^{\lambda_q\Psi}\ast F)=0$
that a homogeneous perturbation 
of the form field decouples from other fluctuations, while
it couples for inhomogeneous modes\cite{Bousso}.

We again perform the conformal translation to put the action
into canonical form:
%
\begin{eqnarray}
\bar{g}_{\mu\nu}=e^{\frac{2q}{p-2}\phi}g_{\mu\nu}.
\end{eqnarray}
Then we have the $p$-dimensional effective action
%
\begin{eqnarray}
S&=&\frac{1}{16\pi}\int d^p x \sqrt{-\bar{g}}
\biggl[{}^{(p)}\bar{R}
-\frac12 D_{\mu}\Psi D^{\mu}\Psi
\nonumber
\\
&&\;\;\;\;\;
-\frac{q(D-2)}{p-2}D_{\mu}\phi D^{\mu}\phi-U(\Psi,\phi)
\biggr],
\end{eqnarray}
where the potential is
%
\begin{eqnarray}
&&U(\Psi,\phi)
=e^{-\frac{2q}{p-2}\phi}
\nonumber
\\
&&\;\;\;\;\;\;\;\;\;
\times\biggl[
-\frac{q(q-1)}{\ell_q^2}e^{-2\phi}
+2e^{\lambda_c \Psi}\Lambda
+\frac{c^2}{2}e^{\lambda_q \Psi}e^{-2q \phi}
\biggr].
\label{potu}
\end{eqnarray}

{}From now on we investigate the stability of the
FR-type solutions.
We showed that the $p$-dimensional background
spacetime is always adS.
It should be noted that the adS spacetime can be
stable for some cases even in the presence of
a tachyonic scalar field.
We start from the positive mass
theorem in adS spacetime\cite{Townsend,PET}.

Let us begin with the derivation of
Breitenlohner and Freedman's stability condition using the
positive energy theorem.
If the potential of the scalar fields can
be written in the form
%
\begin{equation}
U = 2(p-2)^2G^{XY}\frac{\delta W}{\delta \Phi^X}
\frac{\delta W}{\delta \Phi^Y}
 -2(p-1)(p-2)W^2,
\label{super}
\end{equation}
%
the positivity of the total energy is guaranteed
in the Einstein
frame\cite{Townsend}. Indeed, the total energy can be expressed as
%
\begin{eqnarray}
M_{\rm AD}=\int_\Sigma d^{p-1}x
\biggl[2 |\hat D \epsilon|^2 + \sum_X |\delta
 \lambda^{(X)}|^2
\biggr],
\label{energy}
\end{eqnarray}
%
where $\Sigma$ is the asymptotically ${\rm AdS}_p$ space-like
hypersurface,
$\hat D_\mu :=D_\mu +iW (\Phi) \gamma_\mu$ with the spinor covariant
derivative $D_\mu$, and $\epsilon$ is the spinor satisfying the
Witten equation,\footnote{The existence of this solution
should be able to be confirmed by methods similar to those in
Ref.~\cite{Gibbons}}
$\sum_{a=1}^{p-1}\gamma^a \hat D_a \epsilon =0$.
The spinor $\delta \lambda^{(X)}$ is defined by
%
\begin{equation}
\delta \lambda^{(X)}:=\frac1{\sqrt{2}}
\left[i f^{(X)}_Y \gamma^\mu \partial_\mu
\Phi^Y
+2(p-2)f^{(X)Y}\frac{\delta W}{\delta \Phi^Y} \right]\epsilon,
\end{equation}
%
where $f^{(X)}_Y$ is such that
%
\begin{equation}
G_{XY}=:\sum_Zf^{(Z)}_Xf^{(Z)}_Y
\end{equation}
%
holds, which can be thought of as the orthonormal basis
of the target space.

In the perturbative approach, we expand $U$ and $W$ as
%
\begin{eqnarray}
U&=&U_0+\frac{1}{2}\sum_X m_X^2(\bar{\phi}^X)^2,
\\
W&=&W_0+\sum_X W_1^X{\bar{\phi}^X}+\frac12\sum_X
(W_2^X)^2(\bar{\phi}^X)^2.
\end{eqnarray}
%
Here we have diagonalized the target space metric.
{}From the zeroth order equation of  Eq.~(\ref{super}), we find
%
\begin{eqnarray}
W_0^2&=&\frac{1}{2\ell_p^2},
\\
W_1^X&=&0.
\end{eqnarray}
%
Here we have used $U_0=-(p-1)(p-2)/\ell_p^2$.
The second order equation gives
%
\begin{eqnarray}
&&\sum_X\left[4(p-2)^2(W_2^X)^2-2(p-1)(p-2)W_0W_2^X
\right. \nonumber
\\
&& \;\;\;\;\;\;\;\;\;\;\;\;\;
\left.-\frac{1}{2} m_X^2\right]
(\bar{\phi}^X)^2=0.
\end{eqnarray}
%
The condition that a real root of $W_2^X$ exists  is
%
\begin{eqnarray}
(p-1)^2(p-2)^2W_0^2+2(p-2)^2m_X^2\geq0,
\end{eqnarray}
%
which is rewritten in the form
%
\begin{eqnarray}
m_X^2 \ell_p^2\geq -\frac{(p-1)^2}{4}.
\label{BFbound}
\end{eqnarray}
%
This is just the stability bound obtained by
Breitenlohner and Friedmann in
4-dimensional adS spacetime\cite{BF}
and extended to general dimension in
\cite{Mezincescu,Brein,Townsend}.

The potential (\ref{potu}) is expanded around the
background solution
$\phi=0$, $\Psi=\Psi_c$ as
%
\begin{eqnarray}
U(\Psi,\;\bar{\phi}) & = & U(\Psi_c,\;0)
+\frac{1}{2}m_{\Psi}^2(\Psi -\Psi_c)^2
\nonumber \\
& & +m^2 (\Psi - \Psi_c) \bar{\phi}
+\frac{1}{2}m_{\bar{\phi}}^2\bar{\phi}^2,
\end{eqnarray}
where
%
\begin{eqnarray}
m_{\Psi}^2 &=& \frac{p-q-2}{2(p-q)}\lambda_q^2c^2e^{\lambda_q\Psi_c},
\\
m_{\bar{\phi}}^2
&=&\frac{1}{2(D-2)}\left[(2q-1)(p-2)+q\right]c^2e^{\lambda_q\Psi_c},
\\
m^2  &=&
-\sqrt{\frac{q(p-2)}{2(D-2)}}\lambda_qc^2e^{\lambda_q\Psi_c}.
\end{eqnarray}
We can observe that
%
\begin{eqnarray}
m_{\Psi}^2 =m^2 = 0,
\;\;\;
m_{\bar{\phi}}^2& = & \frac{p-1}{2}c^2,
\end{eqnarray}
for $p=q$.

The linear perturbation equation is
%
\begin{eqnarray}
D_{\mu}D^{\mu}{\Psi\choose \bar{\phi}}=M{\Psi\choose \bar{\phi}},
\end{eqnarray}
where the mass matrix is
%
\begin{eqnarray}
M=
\left(
\begin{array}{cc}
m_{\Psi}^2 & m^2
\\
m^2 & m_{\bar{\phi}}^2
\end{array}
\right).
\end{eqnarray}
This equation is diagonalized as
%
\begin{eqnarray}
\label{eq2}
D_{\mu}D^{\mu}\chi_{\pm}=m_{\pm}^2\chi_{\pm},
\end{eqnarray}
where
%
\begin{eqnarray}
\label{mass2}
m_{\pm}^2=\frac{e^{\lambda_q \Psi_c}c^2}{2}
\left[B\pm\sqrt{B^2+\frac{D-2}{p-q}\lambda_q^2}
\right],
\end{eqnarray}
%
\begin{eqnarray}
A&=&(2q-1)(p-2)+q>q,
\\
B&=&\frac{1}{2}\left[\frac{p-q-2}{p-q}\lambda_q^2
+\frac{A}{(D-2)}\right].
\end{eqnarray}

In the present case, the stability bound (\ref{BFbound})
is expressed as
%
\begin{eqnarray}
F_1(p,q,a):=\frac{p-1}{2}+B
-\sqrt{B^2+\frac{D-2}{p-q}\lambda_q^2  } \geq 0.
\label{cri1}
\end{eqnarray}
It should be empasized that this condition is independent of
the cosmological constant
$\Lambda$. This is in contrast with the non-dilatonic case where the
criterion strongly depends on $\Lambda$\cite{Bousso}.

Since our model has three parameter $p$, $q$ and $a$,
it is difficult to express the
condition for stability explicitly
in the general case. So let us focus on some specific cases.

Since the perturbation equation (\ref{eq2}) has always
one tachyonic mode for $p>q$ 
($\Lambda < 0$, $\lambda_c \lambda_q >0$),
the background solution corresponds to the saddle point of the
potential.
On the other hand, for $p\leq q$ ($\lambda_c \lambda_q \leq 0$), $B$ is 
positive semidefinite and
$m_{\pm}^2\geq0$. Hence the adS vacuum is stable independently
of the dilaton coupling constant
in this case.
When we consider higher-dimensional theories with $D\geq 10$,
such as superstring/M theories, our result shows that
Freund-Rubin compactification gives us stable
4- or 5-dimensional spacetime.

One might be interested in the case $a=1$. In this case
$\lambda_q= (p-q)/\sqrt{2(D-2)}$ and
$B=(D-2)/4>0$. The condition for the stability becomes
%
\begin{eqnarray}
2p^2+(q-7)p+q+3 \geq 0.
\end{eqnarray}
%
It is easy to see that this holds for $p \geq 3$. So
we can conclude
that the adS vacuum is stable.

For general $a$, however, there are parameter windows for
the unstable solutions. This can easily be seen for
$a \simeq (D-2)/(D-1)$. In this case $\lambda_q  \gg 1$ and
%
\begin{equation}
F_1 \simeq
\frac{p-1}{2}-\frac{D-2}{p-q-2}.
\end{equation}
%
For example, for $p-2 =q+1$, $F_1 \simeq -(3/2)q <0$.
Thus, this case is definitely unstable. 


Let us summarize the present work. We considered the effect
of the dilation field
on FR compactification with a cosmological constant. As a
result, Minkowski or de Sitter background spacetimes cannot be
realized, while the only solution compatible with the dilaton
field is adS spacetime. We also addressed the latter's
stability. We know that the adS spacetime obtained from FR
compactification in the non-dilatonic case is stable. On the other
hand, as expected, we find an unstable background solution if the
dilaton coupling exists. For the string coupling case($a=1$), the 
solution is stable at low energy.

We have considered several specific cases in the low-energy 
limit (homogeneous perturbation). Hence it should be noted that 
our stability analysis is not definitive. Especially it is reported 
that instability does not come from the homogeneous mode but
inhomogeneous modes in the non-dilatonic cases\cite{Bousso}.
A more systematic study, possibly including non-linear
stability analysis, would be a valuable subject for future study.

\section*{Acknowledgments}

We would like to thank Kei-ichi Maeda and Daisuke Ida
for fruitful discussions. We also give special thanks to 
James Overduin for reading this manuscript carefully. 
TS's work is supported by Grant-in-Aid for Scientific
Research from Ministry of Education, Science, Sports and Culture of
Japan(No. 13135208, No.14740155 and No.14102004).




\begin{references}
\bibitem{quanta}
The different type of the spontaneous compactification
by the quantum effect is also discussed in type IIB
superstring theory by K. Uzawa and K. Maeda.
Private communications.
\end{references}
\end{document}